\documentclass[aps,twocolumn,showpacs,pre]{revtex4-1}
\usepackage{graphicx,graphics}
\usepackage{amsmath,amssymb,amsfonts}
\usepackage{multirow}
\usepackage{hyperref}
\usepackage{graphicx}
\usepackage{color}

\begin{document}
\title{\Large Nonlocal biased random walks and fractional transport on directed networks}
\author{A.P. Riascos}
\email{aperezr@fisica.unam.mx}
\affiliation{Instituto de F\'isica, Universidad Nacional Aut\'onoma de M\'exico, 
	Apartado Postal 20-364, 01000 Ciudad de M\'exico, M\'exico}
\author{T.M. Michelitsch} 
\affiliation{Sorbonne Universit\'e, Institut Jean le Rond d'Alembert, CNRS UMR 7190,4 place Jussieu, 75252 Paris cedex 05, France}
\author{A. Pizarro-Medina} 
\affiliation{Departamento de F\'isica, Facultad de Ciencias, Universidad Nacional Aut\'onoma de M\'exico, Ciudad Universitaria,  Ciudad de M\'exico 04510, M\'exico}
\date{\today}
\begin{abstract}
In this paper, we study nonlocal random walk strategies generated with the fractional Laplacian matrix of directed networks. We present a general approach to analyzing these strategies by defining the dynamics as a discrete-time Markovian process with transition probabilities between nodes expressed in terms of powers of the Laplacian matrix. We analyze the elements of the transition matrices and their respective eigenvalues and eigenvectors, the mean first passage times and global times to characterize the random walk strategies. We apply this approach to the study of particular local and nonlocal ergodic random walks on different directed networks; we explore circulant networks, the biased transport on rings and the dynamics on random networks. We study the efficiency of a fractional random walker with bias on these structures. Effects of 
ergodicity loss which occur when a directed network is not any more strongly connected are also discussed.
\end{abstract}

\maketitle

\section{Introduction}
The study and understanding of dynamical processes taking place on networks have a significant impact in science and engineering with important applications in physics, biology, social and computer systems among many others \cite{VespiBook,MasudaPhysRep2017}. In particular, the diffusion problem associated to the dynamics of a random walker that hops visiting the nodes of the network following different strategies is an important and challenging field of research due to connections with  interdisciplinary topics like ranking and searching on the web \cite{Brin1998,LeskovecBook2014,ShepelyanskyRevModPhys2015}, aging and accumulation of damage \cite{RiascosWangMichelitsch_PRE2019}, the understanding of human mobility in urban settlements \cite{RamascoGeogr2016,Barbosa2018,RiascosMateosPlosOne2017,LoaizaMonsalvePlosOne2019, RiascosMateosSciRep2020}, epidemic spreading \cite{BrockmannPRX2011,ValdezBraunsteinHavlin2020}, algorithms for extracting useful information from data \cite{BlanchardBook2011}, just to mention a few examples. Several types of random walk strategies on networks have been introduced in the last decades, some of them only require local information of  each node and in this way, the walker moves from one node to one of its nearest neighbors \cite{Montroll1965,HughesBook,NohRieger}, whereas in other cases, the total architecture of the network comes into play and nonlocal strategies can use all this information to define long-range hops between distant nodes \cite{RiascosMateos2012,Estrada2017Multihopper}.
\\[2mm]
In addition to the nonlocal random walks mentioned before, we have the fractional diffusion on undirected networks \cite{RiascosMateosFD2014,RiascosMateosFD2015,RiascosMateosQM2015, Michelitsch2016Chaos,Michelitsch2017PhysA,deNigris2017,Michelitsch2017PhysARecurrence, RiascosMichelitsch2017_gL,FractionalBook2019,Perkins2019}, a process associated with a L\'evy like dynamics where the transition probabilities between nodes are defined in terms of powers of the Laplacian matrix of the network \cite{RiascosMateosFD2014,RiascosMichelitsch2017_gL,FractionalBook2019}. This mechanism to generate nonlocality combines the information of all possible paths connecting two nodes on the network \cite{RiascosMateosFD2015} improving the capacity to visit the nodes, a result that offers a significant advantage in networks with large average distances between nodes like lattices and trees but that also is evident in small-world networks \cite{RiascosMateosFD2014}. Beyond the study of nonlocal dynamical processes on networks; recently, the concept of fractional Laplacian of a network has been implemented in semi-supervised learning algorithms for the classification of data structures \cite{Bautista2019}. Other potential applications of nonlocal dynamics on networks require the extension of all this formalism to the case of directed weighted networks \cite{benzi2020fractional}.
\\[2mm]
It is important to notice that in a {\it connected undirected network} and in a {\it strongly connected directed graph} the fractional Laplacian matrix (i.e. the matrix function that is generated by fractional powers of the Laplacian matrix) generates a {\it fully connected topology} corresponding to a network with connections between all nodes of the network
with characteristic asymptotic power-law decay. 
An undirected graph is referred to as `{\it connected}' if between any pair of nodes exists a path of finite length. A directed graph is called `{\it strongly connected}' if for any pair of nodes $(ij)$ there are directed paths $i \rightarrow j$ and $j\rightarrow i$, or in other words any node can be reached from any other node by a finite number of steps (see Ref. \cite{benzi2020fractional} for definitions and outline of properties). Connected undirected and strongly connected directed graphs
fulfill the condition of {\it aperiodic ergodicity}. Conversely aperiodic ergodic graphs always are either connected undirected networks or strongly connected directed graphs.
\\[2mm]
The fractional Laplacian contains the complete information on the topology of the network. For an outline how long-range interactions of asymptotic power-law decay in harmonic systems modify their spectral properties and universal features, we refer to \cite{BurioniRevModPhys1997}. The spectral dimension of general networks is analyzed in the seminal paper \cite{HatoriWatanabe1987} and applications of this approach in spin models on graphs is outlined in the article \cite{BurioniCassiVezzani1999}, the Laplacian spectrum of simplicial complexes using the renormalization group is discussed in Ref. \cite{Bianconi_2020}.
For a general analysis of the spectral dimension of (unbiased) L\'evy flights in the $\mathbb{R}^d$ we invite the reader to consult Chapter 8 in Ref. \cite{FractionalBook2019}. 
\\[2mm]
In the present paper, we explore the dynamics of a random walker with transition probabilities defined in terms of the elements of the fractional Laplacian matrix in directed networks. In the first part, general definitions and properties of the fractional transport, ergodicity, and emergence of nonlocality on strongly connected directed networks are discussed. The formalism generalizes different results and techniques developed in the context of the fractional Laplacian of an undirected graph \cite{FractionalBook2019}. In contrast to the unbiased transport defined through symmetric weighted matrices \cite{riascos2019random}, in the directed case the eigenvalues of fractional transition matrices are complex numbers. We explore different types of directed structures, especially circulant directed networks such as directed rings, but also random directed networks of the Erd\H{o}s-R\'enyi type. In the case of directed rings, we show analytically the connection of fractional dynamics with nonlocal random walks similar to L\'evy flights where the property of strong connectivity generates aperiodic ergodicity in the fractional walk. We demonstrate that in directed networks which are not strongly connected the fractional Laplacian has zero matrix elements, and
conversely if the fractional Laplacian matrix has uniquely non-zero entries, the directed graph is 
strongly connected where the resulting walk is aperiodic ergodic.
\\[2mm]
We also analyze the efficiency of fractional random walk strategies that emerge in directed networks using mean-first passage times and global times that characterize the transport. The implementation of these measures allows identifying cases where the combination of biased transport and nonlocality is an inefficient strategy to explore a network; this particular result is in contrast with the efficiency on undirected networks for which the fractional dynamics always improve the speed of the exploration in comparison with a local random walker \cite{FractionalBook2019}. The general approach introduced reveals several cases where the combination of nonlocal displacements and the bias generated by the directions of lines produce a global effect that can either reduce or improve the efficiency of a random walker to visit all the nodes or reach a particular target on the network.
\section{Fractional Laplacian of directed networks}
In this section, we introduce a generalization of the fractional Laplacian of undirected networks (see Refs. \cite{RiascosMateosFD2014,RiascosMichelitsch2017_gL,FractionalBook2019}) to a general class of directed weighted networks. In terms of this operator, we define transition probabilities of a Markovian random walker associated to the biased fractional transport on networks.
\\[3mm]
We consider directed weighted networks with $N$ nodes $i=1,\ldots ,N$. The topology of the network is described by an adjacency matrix $\mathbf{A}$ with elements $A_{ij}=1$ if there is an edge between the nodes $i$ and $j$ and $A_{ij}=0$ otherwise. In addition to the network structure, we have a $N\times N$ matrix of weights $\mathbf{\Omega}$ with elements $\Omega_{ij}\geq 0$. The matrix $\mathbf{\Omega}$ can include information of the structure of the network or incorporate additional data describing the flow capacity of each link \cite{Flow_graphsPRE2011,riascos2019random,RiascosWangMichelitsch_PRE2019}. In the simplest case,  $\mathbf{\Omega}$ coincides with the adjacency matrix $\mathbf{A}$. 
\\[2mm]
Since the matrix of weights in general is not symmetric, we define two types of degrees associated to each node. First, we have the {\it in-degree} given by
\begin{equation}\label{kin_deg}
k_i^{(\mathrm{in})}=\sum_{l=1}^N \Omega_{li}.
\end{equation}
This degree determines the total flow to reach the node $i$ from all the nodes. In a similar way, we have the {\it out-degree}  
\begin{equation}\label{kout}
k_i^{(\mathrm{out})}=\sum_{l=1}^N \Omega_{il},
\end{equation}
that quantifies the total flow from the node $i$ to all the nodes in the network. Without loss in the generality of the formalism, we assume also that $\Omega_{ii}=0$ for $i=1,2,\ldots,N$. In the following, we consider connected directed networks for which $k_i^{(\mathrm{out})}>0$ for all the nodes. 
\\[2mm]
\begin{figure*}[!t]
\begin{center}
\includegraphics*[width=0.9\textwidth]{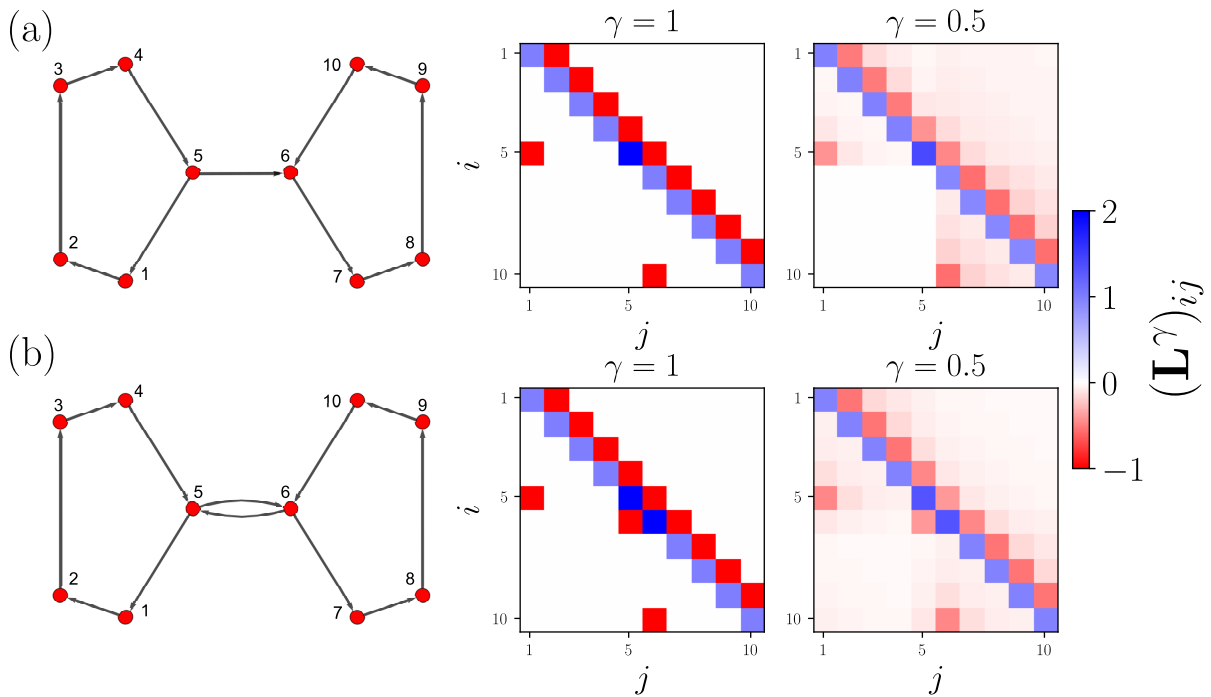} 
\end{center}
\vspace{-5mm}
\caption{\label{Fig_1} (Color online) Fractional Laplacian of directed networks with $N=10$ nodes. Graphs formed with two directed rings and (a) a directed edge from $5$ to $6$ and (b) a symmetric link connecting these two nodes. For each network we present the Laplacian matrix (local limit with $\gamma=1$) and the fractional Laplacian $\mathbf{L}^\gamma$ with $\gamma=0.5$, the values of entries $i$, $j$ are codified in the colorbar. For the fractional Laplacian $\gamma=0.5$, we evaluate numerically Eq. (\ref{LfracDef}).}
\end{figure*}
In terms of the matrix of weights, we define the Laplacian matrix $\mathbf{L}$ with elements $i$, $j$, given by
\begin{equation}\label{Laplacian}
L_{ij}=k_i^{(\mathrm{out})} \delta_{ij}-\Omega_{ij}
\end{equation}
where $\delta_{ij}$ denotes the Kronecker's delta. Equation (\ref{Laplacian}) is a generalization of the Laplacian matrix for binary undirected networks \cite{NewmanBook,Mohar1991lsg,Mohar1997sal}, to include the possibility of weights in the connections and asymmetry in the flow on some lines, for these particular connections $\Omega_{ij}\neq \Omega_{ji}$. It is noted that dynamical processes in directed networks have a greater variety than in the undirected case. For example, for the matrix $\mathbf{L}$, the diagonal could be defined in terms of the in-degree in Eq. (\ref{kin_deg}) producing a different process. Our choice in Eq. (\ref{Laplacian}) is motivated by diffusive transport and the effect of nonlocality, similar nonlocal effects have been found in human mobility in different types of transport described by directed networks \cite{RiascosMateosPlosOne2017,LoaizaMonsalvePlosOne2019,RiascosMateosSciRep2020}.
\\[3mm]
On the other hand, in the context of the fractional diffusion on networks is introduced the fractional Laplacian  matrix $\mathbf{L}^{\gamma}$, where $\gamma$ is a real number ($0<\gamma<1$). The resulting operator models the fractional dynamics on general networks \cite{RiascosMateosFD2014,RiascosMichelitsch2017_gL,FractionalBook2019}. Using Dirac's notation for the eigenvectors, we have a set of right eigenvectors $\{\left|\Psi_j\right\rangle\}_{j=1}^N$  that satisfy the eigenvalue equation 
$\mathbf{L}\left|\Psi_j\right\rangle=\mu_j\left|\Psi_j\right\rangle$ for $j=1,\ldots,N$. With this information, we define the matrix $\mathbf{Q}$  with elements $Q_{ij}=\left\langle i|\Psi_j\right\rangle$  and the diagonal matrix $\mathbf{\Lambda}=\textrm{diag}(\mu_1,\mu_2,\ldots,\mu_N)$. These matrices satisfy $\mathbf{L}\,\mathbf{Q}=\mathbf{Q}\,\mathbf{\Lambda}$, therefore
\begin{equation}
\mathbf{L}=\mathbf{Q}\mathbf{\Lambda}\mathbf{Q}^{-1},
\end{equation}
where $\mathbf{Q}^{-1}$ is the inverse of $\mathbf{Q}$. Using the matrix $\mathbf{Q}^{-1}$, we define the set of left eigenvectors $\{\left\langle \bar{\Psi}_i\right|\}_{i=1}^N$ with components $\left\langle \bar{\Psi}_i|j\right\rangle=(\mathbf{Q}^{-1})_{ij}$. Therefore
\begin{equation}\label{LfracDef}
	\mathbf{L}^{\gamma}=\mathbf{Q} \mathbf{\Lambda}^{\gamma} \mathbf{Q}^{-1}
	=\sum_{m=1}^N \mu_m^{\gamma}\left|\Psi_m\right\rangle\left\langle \bar{\Psi}_m\right| ,
\end{equation}
where $\mathbf{\Lambda}^{\gamma}=\textrm{diag}(\mu_1^\gamma,\mu_2^{\gamma},\ldots,\mu_N^{\gamma})$ for $0<\gamma\leq 1$. It is sufficient here for our aims to consider uniquely the diagonalizable cases. For an outline where the Laplacian has Jordan canonical form we refer to the article \cite{benzi2020fractional}.
\\[2mm]
In Fig. \ref{Fig_1}, we show the fractional Laplacian $\mathbf{L}^{\gamma}$ for directed networks with $N=10$. In this case, we choose $\mathbf{\Omega}$ as the adjacency matrix of the network. These graphs are formed by two directed rings (the first ring with nodes $1$ to $5$ and the second one with nodes $6$ to $10$) and a directed line connecting the two rings (see Fig. \ref{Fig_1}(a)) or a symmetric link between nodes $5$ and $6$ (as depicted in Fig. \ref{Fig_1}(b)). The network in Fig. \ref{Fig_1}(a) is not (strongly) connected, see for example that there are no paths starting from one of the nodes $6$ to $10$ ending in nodes $1$ to $5$, in these cases the distances between nodes are infinite. In this directed graph, we observe that the respective elements of the fractional Laplacian are also null (i.e. the block $(\mathbf{L}^{\gamma})_{rl}=0$ for $r=6,\ldots,10$ and $l=1,\ldots ,5$), this is a consequence of the fact that $(\mathbf{L}^\gamma)_{ij}$ incorporates information of all the possible paths connecting $i$ with $j$, then when distance $d_{ij}\to \infty$, $(\mathbf{L}^\gamma)_{ij}=0$ for $0<\gamma\leq 1$ (see Ref. \cite{FractionalBook2019}). In contrast, the structure in Fig. \ref{Fig_1}(b) is strongly connected, hence, there is a path of finite length connecting any pair of nodes of the network, in this case, we see that all the elements of $(\mathbf{L}^\gamma)_{ij}$ are non-null, drawn for $\gamma=0.5$, as a result of the full connectivity of that network.
\\[2mm]
Back to the general case with strongly connected networks, due to the asymmetry of $\mathbf{\Omega}$, the eigenvalues $\mu_l$ can take complex values. However, as a consequence of $\sum_{l=1}^N L_{il}=0$, the definition of the out-degree in Eq. (\ref{kout}) and the conditions $L_{ii}>0$,  $L_{ij}\leq 0$ for $i\neq j$, the fractional Laplacian $\mathbf{L}^\gamma$ of a directed weighted network has real entries and fulfills the following properties for $0<\gamma\leq 1$:
\\[2mm]
{\bf (i)} For the {\it fractional out-degree}, we have
\begin{equation}
\label{fracout}
 k_i^{(\gamma)}\equiv(\mathbf{L}^\gamma)_{ii}=
 -\sum_{m\neq i} (\mathbf{L}^\gamma)_{im}.   
\end{equation}
Condition (i) reflects the property that the zero eigenvalue of the Laplacian matrix is conserved by the fractional Laplacian (where the corresponding eigenvector has constant components).
\\[2mm]
{\bf (ii)} The diagonal elements of $\mathbf{L}^\gamma$ are positive real values; in this way  $k_i^{(\gamma)}>0$ for $i=1,2,\ldots, N$. 
\\[2mm]
{\bf (iii)} The non-diagonal elements of  $\mathbf{L}^\gamma$ are real values
satisfying $(\mathbf{L}^\gamma)_{ij}\leq 0$ for $i\neq j$. See Refs. \cite{RiascosMichelitsch2017_gL,FractionalBook2019,benzi2020fractional} for a detailed discussion on these properties for undirected and directed networks, respectively. 
\\[2mm]
Considering the following integral representation of the 
fractional Laplacian matrix \cite{ RiascosMichelitsch2017_gL,FractionalBook2019}
\begin{equation}
\label{mellin-repres}
{\mathbf L}^{\gamma} =  -\frac{1}{\Gamma(-\gamma)}\int_0^{\infty} t^{-\gamma-1} \, ({\mathbb I}- e^{-{\mathbf L}t})\, {\rm d}t  ,\hspace{0.5cm} 0<\gamma<1
\end{equation}
($-\Gamma(-\gamma) =\frac{\Gamma(1-\gamma)}{\gamma} >0$)
shows that the properties {\bf (i)}-{\bf (iii)} of ${\mathbf L}$ are conserved in the interval of convergence $\gamma \in (0,1)$ of Eq. (\ref{mellin-repres}) (see Appendix \ref{AppendA} for a brief demonstration). In this relation $\mathbb{I}=(\delta_{ij})$ indicates the $N\times N$ identity matrix and $\Gamma(..)$ stands for the Gamma-function. 
We observe that in the fractional interval $\gamma \in (0,1)$ all matrix elements of the fractional Laplacian in Eq.  (\ref{mellin-repres}) are strictly non-zero 
{\it if and only if the directed network is strongly connected}  which is true for the graph in Fig. \ref{Fig_1}(b), however it is not true for the graph in Fig. \ref{Fig_1}(a) which is not strongly connected where some elements of the fractional Laplacian matrix are zero. The fractional Laplacian of a strongly connected structure for $\gamma \in (0,1)$ fulfills $({\mathbf L}^{\gamma})_{ii}>0 $ and $({\mathbf L}^{\gamma})_{ij} <0$ (for $i\neq j$) where all entries are strictly non-zero.
\\[3mm]
The characteristics of the fractional Laplacian matrix allow to define the fractional diffusion on directed weighted networks as a discrete-time Markovian process determined by a transition matrix $\mathbf{W}^{(\gamma)}$ with elements $w_{i\to j}^{(\gamma)}$ representing the probability to hop from $i$ to $j$ given by 
\begin{equation}\label{wijfrac}
w_{i\to j}^{(\gamma)}=\delta_{ij}-\frac{(\mathbf {L}^\gamma)_{ij}}{k_i^{(\gamma)}}\qquad 0<\gamma\leq 1.
\end{equation}
Above properties (i)-(iii) of the fractional Laplacian matrix indeed guarantee stochasticity of the fractional transition matrix (\ref{wijfrac}) in the interval $0<\gamma \leq 1$. Further {\it for a strongly connected directed network} it follows from the above consideration that in the fractional interval $0<\gamma<1$ all off-diagonal elements of the transition matrix $w_{i\to j}^{(\gamma)} >0$ ($\forall i \neq j$) of the fractional walk are strictly positive (with $w_{i\to i}^{(\gamma)}=0$ per construction) and hence $\mathbf{W}^{(\gamma)}$ fulfills the condition of {\it aperiodic ergodicity}.
\\[2mm]
It is worth noticing the role of the nonlocality generated by $\mathbf{L}^\gamma$. If the local random walker ($\gamma=1$) can reach any node in the network in a finite number of steps starting from any node (ergodic condition), $\mathbf{L}^\gamma$ combines the information of all these trajectories in the directed network to define a new nonlocal process that maintains ergodicity. However, this is not the case when the process with $\gamma=1$ is not ergodic as we saw in the example in Fig. \ref{Fig_1} (a).   For this reason, in the next part, we maintain our discussion only for strongly connected weighted networks for which 
$\mathbf{W}^{(\gamma)}$ defines ergodic processes (see Appendix \ref{AppendA} for a proof of aperiodic ergodicity in {\it connected undirected and strongly connected directed graphs}).
\\[3mm]
Here we consider Markovian memoryless walks on directed graphs where at
each time instant $t=0,1,2,\ldots$ the fractional random walker makes a jump from one to another node on the network in a process without memory. The probability  $P_{ij}(t;\gamma)$ to start at time $t=0$ on node $i$ and to reach the node $j$ at time $t$ satisfies the master equation \cite{HughesBook,NohRieger,FractionalBook2019}
\begin{equation}\label{master}
P_{ij}(t+1;\gamma) = \sum_{m=1}^N  P_{im} (t;\gamma) w_{m\rightarrow j} ^{(\gamma)}.
\end{equation}
In the following part, we analyze the consequences of the fractional dynamics defined by Eqs. (\ref{wijfrac})-(\ref{master}) on different directed weighted networks that include circulant directed networks, biased transport on rings and random networks. We explore the characteristics of the eigenvalues of the transition matrix, the probabilities of transition in Eq. (\ref{wijfrac}), mean first passage times and global times that describe the efficiency of the random walker to reach any node on the network.
\section{Fractional random walks on directed circulant networks}
In this section, we explore the fractional transport with transition probabilities given by Eq. (\ref{wijfrac}). We analyze  directed networks defined by matrices of weights $\mathbf{\Omega}$ with a  circulant matrix structure.
\subsection{Circulant matrices}
A circulant matrix $\mathbf{C}$ is an  $n \times n$ matrix with the form \cite{Aldrovandi2001,VanMieghem2011}
\begin{equation} \label{matC}
\mathbf{C}=
\left(
\begin{array}{ccccc}
c_0 & c_{n-1} & c_{n-2} & \ldots & c_1\\
c_1 & c_{0} & c_{n-1} & \ldots & c_2\\
c_2 & c_{1} & c_{0} & \ldots & c_3\\
\vdots & \vdots & \vdots & \ddots & \vdots \\
c_{n-1} & c_{n-2} & c_{n-3} &\ldots & c_0\\
\end{array}\right) \, ,
\end{equation}
with elements $C_{ij}$. Thus, each column has real elements $c_0, c_1,\ldots,c_{n-1}$ ordered in such a way that $ c_0 $ describes the diagonal elements and $C_{ij}=c_{(i-j)\text{mod}\, n}$. In addition to $\mathbf {C}$, the elementary circulating matrix  $\mathbf{E}$ is defined, which has all its null elements except $c_1=1$. From $\mathbf{E}$, the integer powers $\mathbf{E}^l$ for $l=0,1,2,\ldots,n-1$ are also circulant matrices with null elements except $c_l=1$. Therefore, Eq. (\ref{matC}) can be expressed as \cite{VanMieghem2011}
\begin{align}\nonumber
\mathbf{C}&=c_0\mathbb{I}+c_1\mathbf{E}+c_2\mathbf{E}^2+\ldots+c_{n-1}\mathbf{E}^{n-1}\\
&=\sum_{m=0}^{n-1} c_m \mathbf{E}^{m}, \label{matCsum}
\end{align}
where $\mathbb{I}=\mathbf{E}^{0}$ is the  $n\times n$ identity matrix. Furthermore, the relation $\mathbf{E}^n=\mathbb{I}$ requires that the eigenvalues  $\nu$ of $\mathbf{E}$  satisfy $\nu^n=1$; therefore, those eigenvalues are given by \cite{VanMieghem2011}
\begin{equation}
\nu_l=e^{\text{i}\frac{2\pi(l-1)}{n}} \qquad \text{for}\qquad l=1,\ldots, n,
\end{equation}
with $\text{i}\equiv\sqrt{-1}$. The respective eigenvectors $\{|\Psi_m\rangle\}_{m=1}^{n}$ have the components $\langle l |\Psi_m\rangle=\frac{1}{\sqrt{n}}e^{-\text{i}\frac{2\pi}{n}(l-1)(m-1)}$ (see Ref. \cite{VanMieghem2011} for details). Now, using Eq. (\ref{matCsum}), the eigenvectors $|\Psi_l\rangle$ satisfy $\mathbf{C}|\Psi_l\rangle=\eta_l |\Psi_l\rangle$, where the eigenvalues $\eta_l$ are given by \cite{VanMieghem2011}
\begin{equation} \label{SpectCeta}
\eta_l=\sum_{m=0}^{n-1} c_m e^{\text{i}\frac{2\pi}{n}(l-1)\, m}
\end{equation}
for $l=1,2,\ldots, n$. This result defines the eigenvalues of $\mathbf{C}$ in terms of the coefficients $c_0, c_1,\ldots,c_{n-1}$.
\subsection{Directed circulant networks}
\begin{figure}[!t]
\begin{center}
\includegraphics*[width=0.5\textwidth]{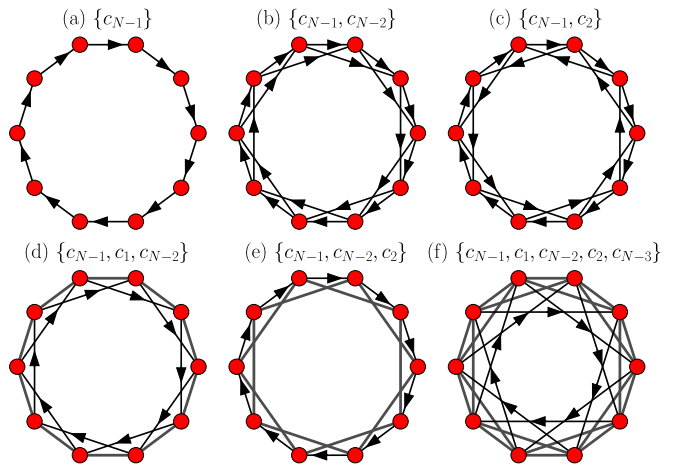} 
\end{center}
\vspace{-5mm}
\caption{\label{Fig_2} (Color online) Circulant directed networks with $N=10$ nodes. We define cyclic structures using circulant adjacency matrices $\mathbf{A}$ with particular non-null elements in Eq. (\ref{matC}). For each network we indicate the coefficients $c_i$ $(i=0,1,\ldots,N-1)$ taking the value 1. Arrows represent the direction of each edge, and connections including both directions are represented with a line.}
\end{figure}
\begin{figure*}[!t]
\begin{center}
\includegraphics*[width=0.9\textwidth]{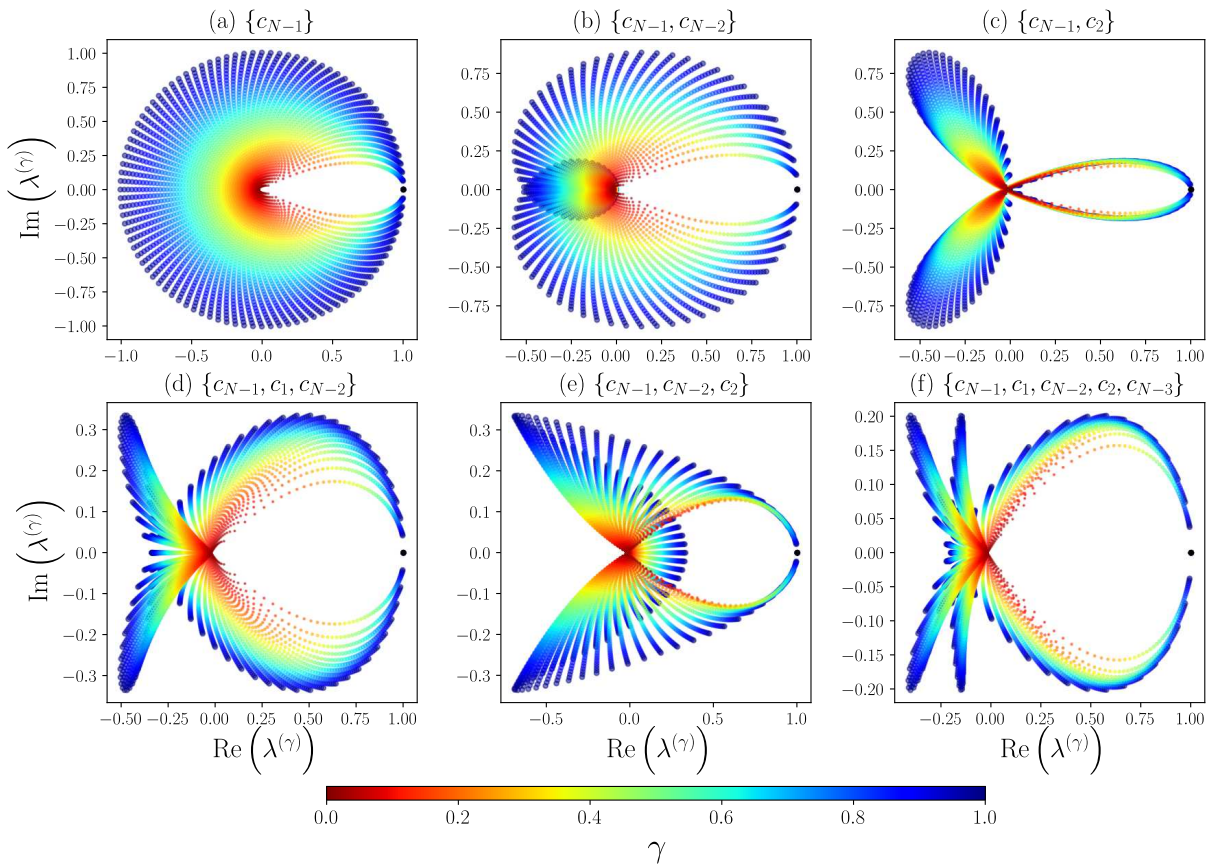} 
\end{center}
\vspace{-7mm}
\caption{\label{Fig_3} (Color online) Eigenvalues of the transition matrix $\mathbf{W}^{(\gamma)}$ for circulant directed networks with $N=100$ represented in the complex plane. Each eigenvalue $\lambda_i^{(\gamma)}$, $i=1,2,\ldots,N$, is determined by Eq. (\ref{lambda_circulant}) with eigenvalues $\mu_l$ given by Eq. (\ref{mu_circulant}). In (a)-(f) we specify the non-null coefficients $c_m=1$ that define the adjacency matrix of the network. For each $\gamma$, we have $N=100$ points and, the effect of the fractional parameter $\gamma$ modifies the set of eigenvalues represented with different colors codified in the colorbar in the interval $0<\gamma\leq 1$. The studied networks  have a similar topology to those presented in Fig. \ref{Fig_2}.}
\end{figure*}
The circulant matrix $\mathbf{C}$ defined in Eq. (\ref{matC}) allows us to explore different directed structures with $N$ nodes and an adjacency matrix $\mathbf{A}$ with specific values $c_1,c_2,\ldots,c_{N-1}$ equal to 0 or 1, the result is a network with a periodic structure. When non-null elements appear in pairs $c_i$ and $c_{N-i}$ $(i=1,2,\ldots,N-1)$ in the adjacency matrix, the structure is an undirected network with symmetric $\mathbf{A}$. Fractional dynamics on undirected circulant networks has been studied in detail for continuous-time random walks \cite{RiascosMateosFD2015,FractionalBook2019}, quantum transport \cite{RiascosMateosQM2015} and diffusion on multilayer networks \cite{Perkins2019}.
\\[2mm]
In the following, we explore cases when the matrix of weights $\mathbf{\Omega}$ that defines the Laplacian in Eq. (\ref{Laplacian}) is  $\mathbf{\Omega}=\mathbf{A}$, this adjacency matrix is not symmetric with a net direction in some lines. For example, the particular network with non-null $c_1=1$ (or $c_{N-1}=1$) produces a directed ring. In Fig. \ref{Fig_2}, we illustrate several directed circulant structures with $N=10$ nodes. In Fig. \ref{Fig_2}(a) we have a directed ring, whereas in Figs. \ref{Fig_2}(b)-(f) other networks are generated by adding new sets of lines defined with non-null elements $c_i$ $(i=1,2,\ldots, N-1)$. In some cases, two nodes are connected with links in both directions. We represent this particular type of connection with lines between nodes as shown in Figs. \ref{Fig_2}(d)-(f).
\\[2mm]
The advantage in the study of the fractional dynamics on circulant networks is that we know all the eigenvalues and eigenvectors of the Laplacian matrix $\mathbf{L}$, since this is also a circulant matrix. In addition, circulant networks are regular with the same fractional out-degree $k^{(\gamma)}=\frac{1}{N}\sum_{m=1}^N \mu_m^\gamma$ for all the nodes. Therefore, the eigenvalues of the transition matrix $\mathbf{W}^{(\gamma)}=\mathbb{I}-\frac{\mathbf{L}^\gamma}{k^{(\gamma)}}$ in Eq. (\ref{wijfrac}) for a fractional random walker are
\begin{equation}\label{lambda_circulant}
\lambda_i^{(\gamma)}=1-\frac{\mu_i^\gamma}{k^{(\gamma)}} \qquad i=1,2,\ldots,N,
\end{equation}
where, applying the result in Eq. (\ref{SpectCeta}) for a network defined by $\mathbf{A}$ with a set of lines with a particular sequence of values 0 and 1 for the coefficients $\{c_m\}_{m=1}^{N-1}$, the eigenvalues of the Laplacian matrix  $\mathbf{L}$ are
\begin{equation}\label{mu_circulant}
\mu_l=c_0-\sum_{m=1}^{N-1} c_m e^{\text{i}\frac{2\pi}{N}(l-1)\, m},
\end{equation}
where the diagonal element $c_0\equiv \sum_{m=1}^{N-1} c_m$ is the out-degree of each node.
\\[2mm]
The result in Eq. (\ref{mu_circulant}) shows that in circulant directed networks the eigenvalues of the Laplacian matrix are complex numbers and, as a consequence, the eigenvalues of $\mathbf{W}^{(\gamma)}$  in Eq. (\ref{lambda_circulant}) are also complex numbers. In Fig. \ref{Fig_3}, we illustrate the effect of the fractional parameter $\gamma$ on the eigenvalues $\lambda_i^{(\gamma)}$ in Eq. (\ref{lambda_circulant}) for different structures with topologies similar to the networks in Fig. \ref{Fig_2}. We show the real and imaginary parts of $\lambda_i^{(\gamma)}$ for directed networks with $N=100$ nodes and $0<\gamma\leq 1$.
\subsection{Long-range dynamics and L\'evy flights}
We have analyzed the spectral properties of circulant matrices associated to the Laplacian $\mathbf{L}$ and fractional transition probabilities $\mathbf{W}^{(\gamma)}$. In the following part, we calculate the probabilities $w_{i \to j}^{(\gamma)}$ and the relation with the distance $d_{ij}$ that gives the shortest-path length between nodes $i$ and $j$. Using Eq. (\ref{mu_circulant}) and the respective eigenvectors of circulant matrices, we have for the fractional Laplacian
\begin{align}\nonumber
(\mathbf{L}^\gamma)_{ij}&=\sum_{l=1}^N \mu_l^\gamma \langle i|\Psi_l\rangle \langle \Psi_l|j\rangle\\
&=\frac{1}{N}\sum_{l=1}^N\mu_l^\gamma e^{\mathrm{i}\frac{2\pi}{N}(l-1)(j-i)}.
\label{Laplij_circulant}
\end{align}
Here, we use the fact that left eigenvectors in circulant matrices satisfy $\left\langle \bar{\Psi}_l\right|=(\left|\Psi_l\right\rangle)^\dagger$, where $\dagger$ denotes the Hermitian conjugate. Therefore, the elements of the fractional transition matrix $\mathbf{W}^{(\gamma)}=\mathbb{I}-\frac{\mathbf{L}^\gamma}{k^{(\gamma)}}$  are
\begin{equation}\label{matWfrac_circ}
w_{i\to j}^{(\gamma)}=\delta_{ij}-\frac{\sum_{l=1}^N \mu_l^\gamma e^{\mathrm{i}\frac{2\pi}{N}(l-1)(j-i)}}{\sum_{l=1}^N \mu_l^\gamma }.
\end{equation}
\begin{figure*}[!t]
	\begin{center}
		\includegraphics*[width=1.0\textwidth]{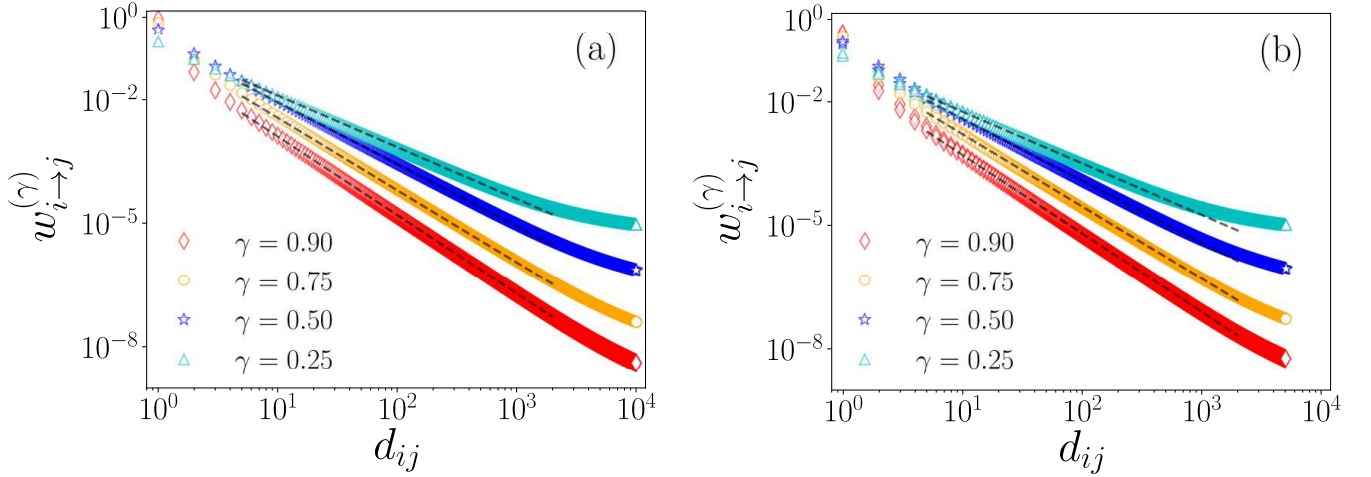}
	\end{center}
	\vspace{-5mm}
	\caption{\label{Fig_4} (Color online) Transition probabilities $w_{i\to j}^{(\gamma)}$ between two nodes as a function of the distance $d_{ij}$ in directed circulant networks with $N=10^4$ nodes. (a) A directed ring defined with $c_{N-1}=1$, (b) a network with $c_{N-1}=c_{N-2}=1$. The probabilities $w_{i\to j}^{(\gamma)}$ are deduced from the analytical relation in Eq. (\ref{matWfrac_circ}) defined in terms of the eigenvalues in Eq. (\ref{mu_circulant}) for $\gamma=0.25,\,0.5,\, 0.75$ and $\gamma=0.9$. Dashed lines represent the inverse power-law relation $w_{i\to j}^{(\gamma)}\propto d_{ij}^{-1-\gamma}$. }
\end{figure*}
With the spectral characteristics of the eigenvalues $\mu_l^\gamma$ and the respective $\lambda_l^{(\gamma)}$ illustrated in Fig. \ref{Fig_3}, the sums in Eq. (\ref{matWfrac_circ}) in the interval $ 0<\gamma\leq 1$ produce well defined probabilities of transition between nodes. In Fig. \ref{Fig_4} we present the transition probabilities $w_{i\to j}^{(\gamma)}$ for circulant networks with $N=10^4$ nodes. Probabilities are presented as a function of the distance $d_{ij}$ connecting the nodes $i$ and $j$ ($d_{ij}$ is the length of the shortest path on the directed structure). In the case of the directed ring [Fig. \ref{Fig_4}(a)], we see the relation $w_{i\to j}^{(\gamma)}\propto d_{ij}^{-1-\gamma}$, a power-law decay also observed in the large-world network explored in Fig. \ref{Fig_4}(b) and exemplified in Fig. \ref{Fig_2}(b). This particular relation between transition probabilities and distances show an emergent dynamics generated through the fractional Laplacian  related with a L\'evy-like dynamics in directed structures. L\'evy flights in undirected networks have been explored in a series of works \cite{RiascosMateos2012,RiascosMateosFD2014,Weng2015,Guo2016,Michelitsch2017PhysA,deNigris2017,Estrada2017Multihopper,FractionalBook2019,Perkins2019,PerkinsPRE2019}, revealing that long-range displacements in undirected networks always improve the capacity to explore a network by inducing dynamically the small-world property \cite{RiascosMateos2012,RiascosMateosFD2014}. Similar long-range strategies have been identified in human mobility \cite{RiascosMateosPlosOne2017}, in the movement of cyclists between stations in bike-sharing systems in Chicago and New York \cite{LoaizaMonsalvePlosOne2019}, in taxi trips in New York City \cite{RiascosMateosSciRep2020} and  in the infection spreading through the United States' highly-connected air travel network \cite{Gustafson2017}.
\subsection{Infinite directed ring}
Now, we explore the fractional Laplacian matrix and transition probabilities in a directed ring. This network is illustrated in Fig. \ref{Fig_2}(a), in the limit $N\to\infty$. In the particular case of a directed ring with $N$ nodes, Eq. (\ref{Laplij_circulant}) takes the form (we use the value $c_{N-1}=1$)
\begin{equation}\label{Lijringdirected}
(\mathbf{L}^\gamma)_{ij}=\frac{1}{N}\sum_{l=1}^N\left(1-e^{-\mathrm{i}\frac{2\pi}{N}(l-1)}\right)^\gamma e^{\mathrm{i}\frac{2\pi}{N}(l-1)(j-i)}.
\end{equation}
However, in the limit of $N$ large, we can define a continuous variable $\varphi=\frac{2\pi}{N}(l-1)$ and $d\varphi=\frac{2\pi}{N}$. Therefore, the elements of the fractional Laplacian matrix for an infinite directed ring are given by
\begin{align}\nonumber
&(\mathbf{L}^\gamma)_{ij}=\frac{1}{2\pi}\int_0^{2\pi}(1-e^{-\mathrm{i}\varphi})^\gamma e^{\mathrm{i}\varphi(j-i)}d\varphi ,\qquad \gamma \in (0,1] \\\nonumber
&=(-1)^{j-i+1}\csc (\pi  \gamma )\frac{\sin [\pi  (j-i-\gamma )] \Gamma (j-i-\gamma)}{\Gamma (-\gamma ) \Gamma (j-i+1)}\\
&=\frac{\Gamma (j-i-\gamma)}{\Gamma (-\gamma ) \Gamma (j-i+1)} ,\hspace{0.5cm} j \geq i.
\label{FracLap_infdir_ring}
\end{align}
\begin{figure*}[!t]
	\begin{center}
		\includegraphics*[width=1.0\textwidth]{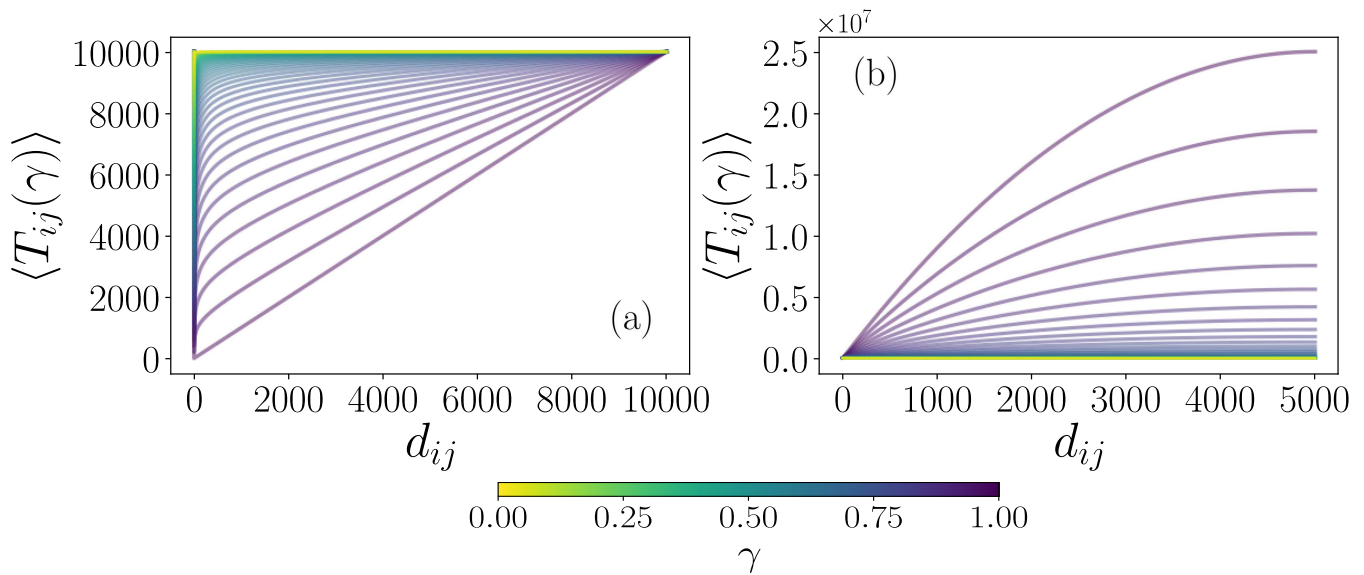} 
	\end{center}
	\vspace{-7mm}
	\caption{\label{Fig_5} (Color online) Mean first passage time  $\left\langle T_{ij} (\gamma)\right\rangle$ as a function of the distance $d_{ij}$ in circulant networks with $N=10^4$ nodes. (a) A directed ring defined with $c_{N-1}=1$,  (b) an undirected ring with  $c_{N-1}=c_{1}=1$. Numerical values are obtained from  Eq. (\ref{TijSpectCirculant}) defined in terms of the eigenvalues in Eq.  (\ref{lambda_circulant}). We codified in the colorbar the different values $0<\gamma\leq 1$. }
\end{figure*}
In particular, $(\mathbf{L}^\gamma)_{ij}=0$ for $j < i$ and $(\mathbf{L}^\gamma)_{jj}=1$ for the diagonal elements. The fractional Laplacian (\ref{FracLap_infdir_ring}) of the infinite ring is an upper triangular matrix where all
entries below the diagonal are null.
The infinite ring of our example hence is (unlike the finite ring) not any more strongly connected and hence not any more ergodic, and the
corresponding fractional walk on the infinite ring allows only jumps into the positive integer-direction (having a triangular transition matrix given in Eq. (\ref{transmatsibuya})).
Indeed this is a strictly increasing walk into the positive integer direction and can be identified with the so-called {\it Sibuya walk} (See Appendix \ref{AppendB} for some derivations and for a profound analysis of properties consult \cite{PachonPolitoRicciuti2020}).

On the other hand, in the limit $j\gg i$, using the relation $\Gamma(n+\alpha)\approx \Gamma(n)n^\alpha$ for $n$ large, we have for Eq. (\ref{FracLap_infdir_ring}) 
\begin{equation}\label{laplacian_infinite_dring}
(\mathbf{L})_{ij}^\gamma\sim \frac{1}{\Gamma (-\gamma )}\frac{1}{(j-i)^{1+\gamma}}\qquad \text{for} \qquad j\gg i 
\end{equation}
where in this relation $\gamma \in (0,1)$.
Mention worthy in this relation is the local limit $\gamma \to 1^{-}$
which becomes ($j-i=x$)
\begin{equation}
    \label{localimit}
    \begin{array}{l} \displaystyle 
\lim_{\gamma \to 1^{-}}(\mathbf{L}^{\gamma})_{ij}  =\lim_{\gamma \to 1^{-}} - \frac{\gamma}{\Gamma(1-\gamma)} x^{-\gamma-1}  \\ \\ \displaystyle 
= \lim_{\gamma \to 1^{-}} \frac{d}{dx} \frac{x^{-\gamma}}{\Gamma(1-\gamma)} = \frac{d}{dx}\delta(x) =0 ,\hspace{0.5cm} x \hspace{0.1cm} {\rm large}.
\end{array}
\end{equation}
The asymptotic result in Eq. (\ref{laplacian_infinite_dring}) shows analytically that the fractional dynamics in the infinite directed ring produces transition probabilities $w_{i\to j}^{(\gamma)}\propto d_{ij}^{-1-\gamma}$ for $j\gg i$ with distances $d_{ij}=j-i$, this relation is also valid for networks with $N$ large but not necessarily infinite (see Appendix \ref{AppendB} for a detailed discussion). The behavior observed differs from the undirected ring, where a similar analysis reveals a L\'evy-like dynamics with $w_{i\to j}^{(\gamma)}\propto d_{ij}^{-1-2\gamma}$ (see Refs. \cite{RiascosMateosFD2014,RiascosMateosFD2015,RiascosMichelitsch2017_gL}), a result that in the general case of undirected $n-$dimensional lattices is $w_{i\to j}^{(\gamma)}\propto d_{ij}^{-n-2\gamma}$ \cite{Michelitsch2017PhysA,Michelitsch2017PhysARecurrence,FractionalBook2019}. 

\subsection{Efficiency of the fractional transport in circulant structures}
In connected circulant networks, each node has the same fractional degree $k^{(\gamma)}$. In addition, if the structure is strongly connected we have only one eigenvalue $\lambda_1^{(\gamma)}=1$ and; in this particular case, the ergodic condition is fulfilled since, for sufficiently large time, the random walker can reach any node of the network independently of the initial node. Therefore, well-known results for the mean first passage time $\left\langle T_{ij}\right\rangle$, which gives the average number of steps of a discrete-time random walker to start in node $i$ and reach for the first time $j$, still apply for circulant directed structures \cite{MasudaPhysRep2017}.
\\[2mm]
In terms of the left and right eigenvectors ($\left\langle\bar{\phi}_l\right|$ and $\left|\phi_l\right\rangle$, respectively) of the transition matrix $\mathbf{W}$ of a Markovian random walker and the associated eigenvalues $\lambda_l$ (we denote $\lambda_1=1$), we have for $i\neq j$ \cite{RiascosMateos2012,RiascosMichelitsch2017_gL,FractionalBook2019,RiascosWangMichelitsch_PRE2019}
\begin{equation}\label{TijSpect}
	\left\langle T_{ij}\right\rangle
	=\sum_{l=2}^N\frac{1}{1-\lambda_l}\frac{\left\langle j|\phi_l\right\rangle \left\langle\bar{\phi}_l|j\right\rangle-\left\langle i|\phi_l\right\rangle \left\langle\bar{\phi}_l|j\right\rangle}{\left\langle j|\phi_1\right\rangle \left\langle\bar{\phi}_1|j\right\rangle}\, ,
\end{equation}
and the mean first return time $\left\langle T_{ii}\right\rangle=(\left\langle i|\phi_1\right\rangle \left\langle\bar{\phi}_1|i\right\rangle)^{-1}$. In addition, in structures with the same fractional degree we have the global time \cite{RiascosMateos2012,RiascosMichelitsch2017_gL,FractionalBook2019,RiascosWangMichelitsch_PRE2019}
\begin{equation}\label{KconstSpect}
	\mathcal{T}=\sum_{l=2}^N \frac{1}{1-\lambda_l}.
\end{equation}
This is the Kemeny's constant that quantifies the capacity of the process to explore the network in regular structures and  only depends on the eigenvalues of the transition matrix $\mathbf{W}$. 
\\[3mm]
In the fractional transport on circulant networks, the eigenvectors of $\mathbf{L}$, the fractional Laplacian  $\mathbf{L}^\gamma$ and the transition probability matrix $\mathbf{W}^{(\gamma)}$ coincide, since all of them are circulant matrices. Hence, $\left\langle i|\phi_l\right\rangle \left\langle\bar{\phi}_l|j\right\rangle=\langle i|\Psi_l\rangle \langle \Psi_l|j\rangle=\frac{1}{N}e^{\mathrm{i}\frac{2\pi}{N}(l-1)(j-i)}$. Then, for the fractional transport in circulant networks, Eq. (\ref{TijSpect}) takes the form ($i\neq j$)
\begin{figure}[!t]
	\begin{center}
		\includegraphics*[width=0.47\textwidth]{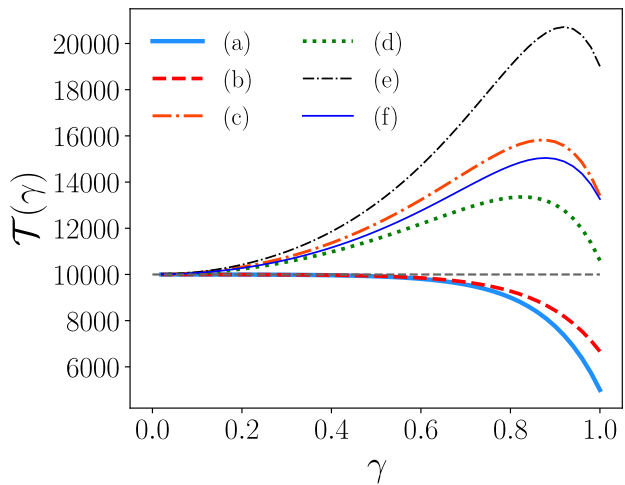} 
	\end{center}
	\vspace{-7mm}
	\caption{\label{Fig_6} (Color online) Kemeny's constant $\mathcal{T}(\gamma)$ as a function of $\gamma$ for different directed networks with $N=10^4$ nodes. The networks have a topology similar to those in Fig. \ref{Fig_2} and are defined with the particular set of coefficients equal to $1$: (a) $\{c_{N-1} \}$, (b) $\{c_{N-1},c_{N-2} \}$, (c) $\{c_{N-1},c_{2} \}$, (d) $\{c_{N-1},c_1,c_{N-2} \}$, (e) $\{c_{N-1},c_{N-2},c_2 \}$, (f) $\{c_{N-1},c_1,c_{N-2},c_2,c_{N-3} \}$.  The results were obtained with Eq. (\ref{KSpectCirculant}) and the Laplacian eigenvalues in Eq. (\ref{mu_circulant}). The dashed horizontal line represents $ \mathcal{T}_{\mathrm{complete}}=(N-1)^2/N$ for a random walker in a complete (fully connected) network.  }
\end{figure}
\begin{equation*}
\left\langle T_{ij} (\gamma)\right\rangle
=\sum_{l=2}^N\frac{1}{1-\lambda_l^{(\gamma)}}\left[1-e^{\mathrm{i}\frac{2\pi}{N}(l-1)(j-i)}\right],
\end{equation*}
and $\left\langle T_{ii} (\gamma)\right\rangle=N$, $\lambda_l^{(\gamma)}$ is given by Eq. (\ref{lambda_circulant}). Therefore
\begin{equation}\label{TijSpectCirculant}
\left\langle T_{ij}(\gamma)\right\rangle
=\left(\frac{1}{N}\sum_{m=2}^N\mu_m^\gamma\right)\sum_{l=2}^N\frac{1-e^{\mathrm{i}\frac{2\pi}{N}(l-1)(j-i)}}{\mu_l^\gamma},
\end{equation}
where we use the result $\mu_1=0$. In a similar way, we obtain for the Kemeny's constant in Eq. (\ref{KconstSpect})
\begin{equation}\label{KSpectCirculant}
\mathcal{T(\gamma)}
=\left(\frac{1}{N}\sum_{m=2}^N\mu_m^\gamma\right)\sum_{l=2}^N\frac{1}{\mu_l^\gamma}.
\end{equation}
Our findings in Eqs. (\ref{TijSpectCirculant})-(\ref{KSpectCirculant}) are valid for directed and undirected connected circulant networks and allow to calculate analytically the mean first passage time (MFPT) and the Kemeny's constant through the specification of the coefficients $c_0,c_1,\ldots,c_{N-1}$ in Eq. (\ref{mu_circulant}).
\\[2mm]
In Fig. \ref{Fig_5}, we depict the MFPT for a directed ring and an undirected ring, both with $N=10^4$ nodes. The results illustrate a completely different behavior in the fractional dynamics in directed and undirected rings. First of all, as we can see in Fig. \ref{Fig_5}(a), for the directed ring defined with an adjacency matrix $c_{N-1}=1$, the results show that for $\gamma=1$ and $i\neq j$,  $\left\langle T_{ij}(\gamma=1)\right\rangle=d_{ij}$, where $d_{ij}=j-i$ for $j\geq i$. In the directed ring, the  value $\gamma=1$ produces a deterministic dynamics where at each step the walker visits a new adjacent node with a cover time $N$. On the other hand, in the interval $0<\gamma<1$, the temporal evolution is stochastic with a biased L\'evy like dynamics increasing the MFPT but maintaining these times below or equal to the value $N$. The results for the biased transport that emerge in the fractional dynamics on the directed ring agree with previous studies showing that L\'evy flights do not always optimize the search problem in the presence of an external drift \cite{PalyulinPNAS2014}. In Fig. \ref{Fig_5}(b), we present the results for times $\left\langle T_{ij}(\gamma)\right\rangle$ in a symmetric ring. In this case, the fractional dynamics with L\'evy flights reduce the MFPT found in the local limit $\gamma=1$ for which the random walker at each step moves with probability $1/2$ from a node to one of its two neighbors.
\\[2mm]
Finally, it is important to have a global time that characterizes the capacity of the random walker to explore the network. In structures such as circulant networks, the Kemeny's constant $\mathcal{T}(\gamma)$ defined in Eq. (\ref{KSpectCirculant}) gives a global value quantifying the efficiency of the fractional random walker to reach all the nodes. In Fig. \ref{Fig_6}, we present the values of $\mathcal{T}(\gamma)$ as a function of $\gamma$ ($0<\gamma\leq 1$) in circulant networks with $N=10^4$ nodes. We explore different directed structures with topologies and properties discussed in Figs. \ref{Fig_2}-\ref{Fig_3}. In Fig. \ref{Fig_6}, the curve (a) describes a fractional random walker in a directed ring with $c_{N-1}=1$.  We can see that the best strategy to explore the network is defined by $\gamma=1$. This is a deterministic limit where the walker visits a new node at each step. With the introduction of L\'evy flights through the fractional dynamics with $0<\gamma<1$, the stochastic transport increases the time $\mathcal{T}(\gamma)$ and in the limit $\gamma=0$ we have $ \mathcal{T}_{\mathrm{complete}}=(N-1)^2/N$ also obtained for a complete network \cite{FractionalBook2019}. In curve (b) we have a structure defined with the elements $c_{N-1}=c_{N-2}=1$, we observe a similar behavior with an optimal value in the local-limit $\gamma\to 1$. In directed structures with more lines like in curves (c)-(f) we see a different behavior where a particular value of $\gamma^\star<1$ produces a maximum in the Kemeny's constant; however, with the reduction of $\gamma$ in the limit $\gamma\to 0$ all the cases approach to the value $ \mathcal{T}_{\mathrm{complete}}$. The results show particular cases where the combination of nonlocal displacements and the bias generated by the directions of the lines produce a global effect that reduces the efficiency to visit all the nodes of the network.
\section{Biased transport on rings}
In the previous section, we studied random walks on circulant directed networks for which we considered the weights $\mathbf{\Omega}=\mathbf{A}$. We are now interested in the effects of the fractional transport when the matrix $\mathbf{\Omega}$ describes some type of bias determined by weights in the links. We explore the transport on a ring with transition probabilities different from the directed and undirected rings studied before.
\\[2mm]
Let us now consider a probability $ 0 \leq \textit{p} \leq 1 $ and a ring with $N$ nodes which are connected only to their first neighbors. In addition, the coefficients $\Omega_{ij}$ are defined by a circulant matrix with non-null elements  $ c_1=p$ and $ c_{N-1}=1-p$. In this way, we  have the probability $p$ to hop in one direction, and $1-p$ to the opposite direction. Using Eq. (\ref{mu_circulant}) for the eigenvalues of the Laplacian matrix, and $c_0=c_1+c_{N-1}=1$, we have
\begin{equation}
\mu_l=1-pe^{\text{i}\varphi_l\, }-(1-p)e^{-\text{i}\varphi_l}
\end{equation}
where $\varphi_l\equiv \frac{2\pi}{N}(l-1)$ and $1\leq l\leq N$.
\begin{figure}[!t]
\begin{center}
\includegraphics[width=0.47\textwidth]{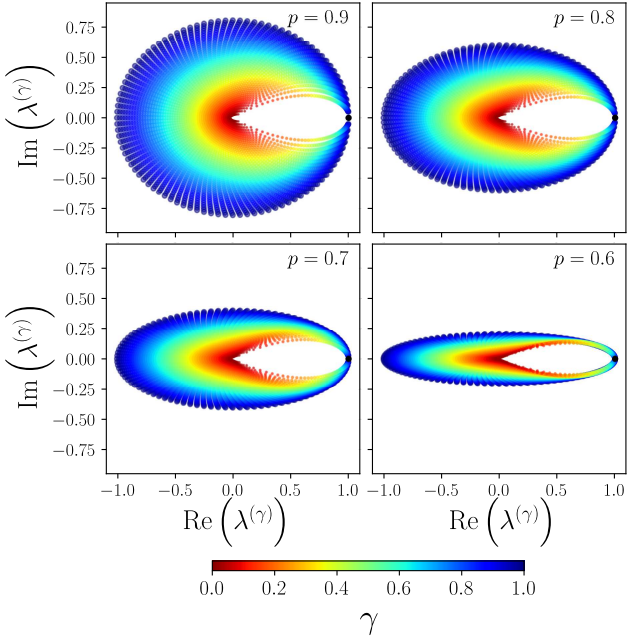}
\end{center}
\vspace{-7mm}
\caption{ \label{Fig_7} (Color online) Eigenvalues of the transition matrix $\textbf{W}^{(\gamma)}$ for biased fractional transport on a ring with $N = 100$ nodes. For $\gamma=1$, the transition probabilities to the two nearest neighbors are $\textit{p}$ to move in one direction and $1-\textit{p}$ in the opposite direction. We explore the effect of $\gamma$ and $p$ in the eigenvalues of $\textbf{W}^{(\gamma)}$ for $\textit{p}$ = {0.9, 0.8, 0.7, 0.6}. Each color represents a different set of eigenvalues, obtained with a given $\gamma$ codified in the colorbar. The eigenvalues are calculated with Eq. (\ref{lambda_biased}). The limit $p=1$ represents a directed ring with eigenvalues analyzed in Fig. \ref{Fig_3}(a).}
\end{figure}

In the general case, the eigenvalues $\lambda_l^{(\gamma)}$  of the fractional transition matrix $\mathbf{W}^{(\gamma)}$ are complex values. By applying Eq. (\ref{lambda_circulant}), we obtain
\begin{equation}\label{lambda_biased}
\lambda_l^{(\gamma)}=1-\frac{1}{k^{(\gamma)}}\left(1-pe^{\text{i}\varphi_l\, }-(1-p)e^{-\text{i}\varphi_l}\right)^\gamma
\end{equation}
with a fractional degree
\begin{equation}
k^{(\gamma)}=\frac{1}{N}\sum_{l=1}^N \left(1-pe^{\text{i}\varphi_l\, }-(1-p)e^{-\text{i}\varphi_l}\right)^\gamma.
\end{equation}
In particular, for $p=1/2$ and $\gamma=1$, we recover the eigenvalues $\lambda_l^{(\gamma=1)}=\cos\left(\varphi_l\right)$ for the local random walk in a symmetric ring. 
\\[2mm]
In Fig. \ref{Fig_7} we show the eigenvalues $\lambda_l^{(\mathrm{\gamma})}$ of the transition matrix $\textbf{W}^{(\gamma)}$ for the biased fractional transport in a ring with $N=100$ nodes. The particular limit $p=1$ recovers the transport on the directed ring presented in Fig. \ref{Fig_3}(a). We see how the bias modeled with the parameter $0.6\leq p \leq 0.9$ reduces the imaginary component of the eigenvalues when $p \to 1/2$. In the limit $p=1/2$ all the eigenvalues are real. We obtain similar results for $0\leq p \leq 0.5$ since, in this interval, the random walker has the same dynamics but with a change in all the directions of the walker.
\begin{figure}[!t]
\begin{center}
\includegraphics[width=0.47\textwidth]{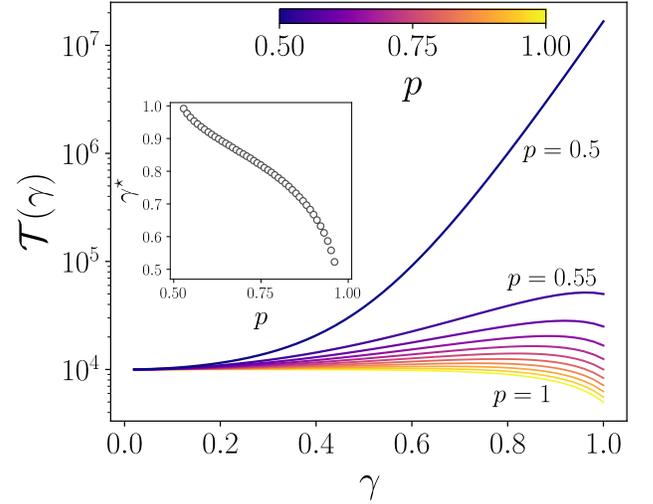}
\end{center}
\vspace{-7mm}
\caption{ \label{Fig_8}  (Color online) Kemeny's constant $\mathcal{T}(\gamma)$ as a function of $\gamma$ for biased fractional transport on a ring with $N=10^4$ nodes. The times $\mathcal{T}(\gamma)$ are calculated with the analytical result in Eq. (\ref{KemenyBiased}). We explore different values of the bias parameter $p$ in the interval $0.5 \leq \textit{p} \leq 1$ codified in the colorbar. The case $p=1/2$ recovers a symmetric random walker with L\'evy flights whereas the limit $p=1$  describes the transport on a directed ring. These two limits are explored in Figs. \ref{Fig_4}-\ref{Fig_5}. In the inset we present the value $\gamma^\star$, as a function of $p$, that maximizes $\mathcal{T}(\gamma)$ and is obtained numerically with Eq. (\ref{relationgammastar}).}
\end{figure}
\\[2mm]
\begin{figure*}[!t]
\begin{center}
\includegraphics[width=0.95\textwidth]{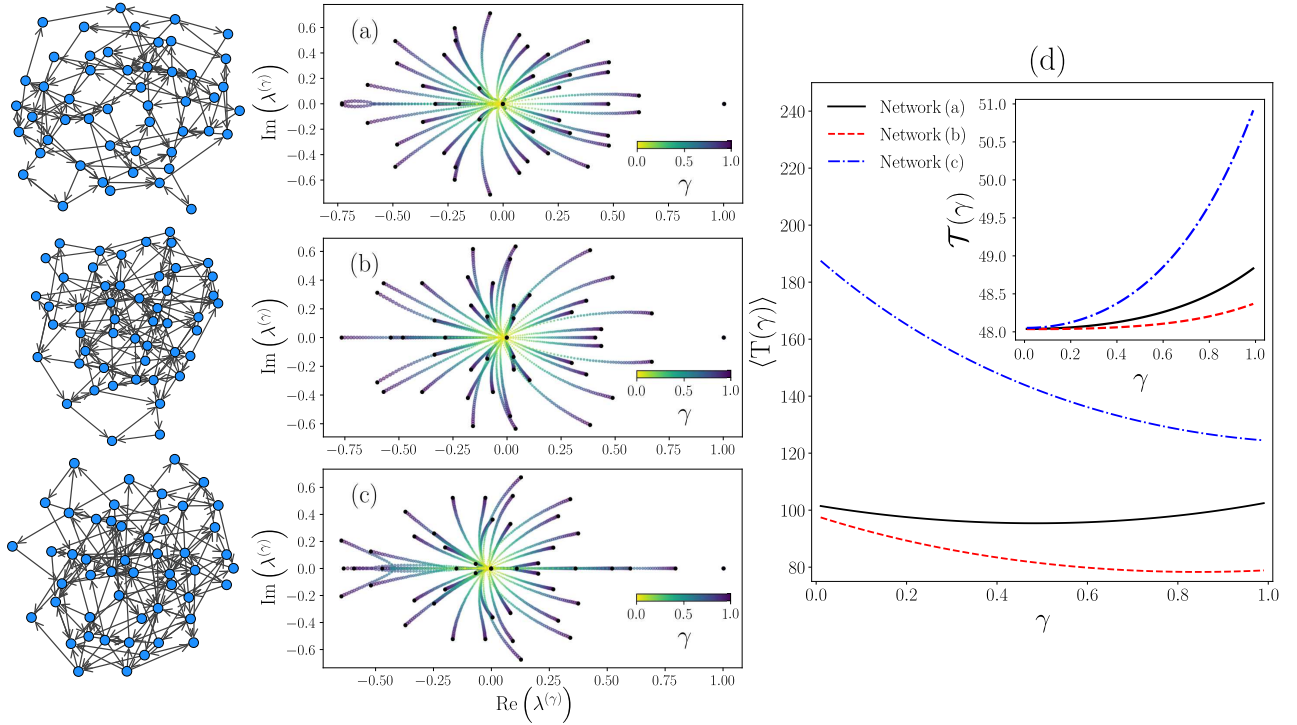}
\end{center}
\vspace{-7mm}
\caption{ \label{Fig_9} (Color online) Fractional transport on directed networks with $N=50$ nodes generated randomly with a probability $p=0.04$ that define the probability to create a connection between two nodes. In (a)-(c) we present three connected networks with the respective eigenvalues of the transition matrix $\mathbf{W}^{(\gamma)}$ in the complex plane. For each $\gamma$ codified in the colorbar in the interval $0<\gamma\leq 1$, we obtain $N=50$ points representing the eigenvalues of $\mathbf{W}^{(\gamma)}$; the eigenvalues for $\gamma=1$ are presented with black dots. In (d) we explore the numerical values of the global time $\langle \mathrm{T}(\gamma)\rangle$ in Eq. (\ref{GlobalMFPT}) for the three networks. The inset shows the results for the Kemeny's constant $\mathcal{T}(\gamma)$.
}
\end{figure*}
Once deduced the eigenvalues of the Laplacian matrix, we can quantify globally the capacity of the fractional random walker to explore the network. Since the fractional degrees are the same for all the nodes, we use the relation for the Kemeny's constant in Eq. (\ref{KSpectCirculant}). Therefore
\begin{equation}\label{KemenyBiased}
\mathcal{T}(\gamma)
=\frac{1}{N}\sum_{l=2}^N\sum_{m=2}^N\left(\frac{1-pe^{\text{i}\varphi_l\, }-(1-p)e^{-\text{i}\varphi_l}}{1-pe^{\text{i}\varphi_m\, }-(1-p)e^{-\text{i}\varphi_m}}\right)^\gamma
\end{equation}
for $0<p\leq 1$ and $0<\gamma\leq 1$. In particular, in the limit $p=1/2$ we recover the result for the symmetric ring 
\begin{equation}
\mathcal{T}_\mathrm{S}(\gamma)=k^{(\gamma)}\sum_{m=2}^N\left(\frac{1}{1-\cos\varphi_m}\right)^\gamma.
\end{equation}
with the fractional degree $k^{(\gamma)}=\frac{1}{N}\sum_{l=2}^N\left(1-\cos\varphi_l\right)^\gamma$. The global time $\mathcal{T}_\mathrm{S}(\gamma)$ is analyzed in detail in Ref. \cite{FractionalBook2019} to characterize the fractional transport on undirected rings.
\\[3mm]
In Fig. \ref{Fig_8} we present the results obtained with Eq. (\ref{KemenyBiased}) for the Kemeny's constant describing the fractional transport with bias in rings for different values of $p$. We observe in the behaviour of the Kemeny's constant that, for $p$ in the interval $0.53\leq p\leq 0.96$, $\mathcal{T}(\gamma)$ presents a maximum for a particular value $\gamma$ denoted as $\gamma^*$. Calculating $\frac{d}{d\gamma}\mathcal{T}(\gamma)\big|_{\gamma=\gamma^\star}=0$, we obtain that the value $\gamma^*$ maximizing the Kemeny's constant satisfies
\begin{equation}\label{relationgammastar}
\sum_{l,m=2}^N\log(z_l)\left(\frac{z_l}{z_m}\right)^{\gamma^\star}=\sum_{l,m=2}^N\log(z_m)\left(\frac{z_l}{z_m}\right)^{\gamma^\star}
\end{equation}
with $z_l\equiv 1-pe^{\text{i}\varphi_l\, }-(1-p)e^{-\text{i}\varphi_l}$. In the inset in Fig. \ref{Fig_8}, we show the values $\gamma ^*$ as a function of $p$,  calculated numerically with  Eq. (\ref{relationgammastar}) . This result determines which L\'evy flight strategy is the least efficient to explore the network, $i.e.$ for which value of $\gamma$ the Kemeny's constant $\mathcal{T}(\gamma)$ has a maximum in the interval $0<\gamma<1$.  
\section{Directed Random Networks}
In this section, we explore the fractional transport on directed networks generated stochastically with an algorithm similar to the Erd\H{o}s-R\'enyi (ER) model \cite{ErdosRenyi1959}. For $N$ nodes we have an adjacency matrix $\mathbf{A}$ and, in each non-diagonal entry $A_{ij}$, we decide to include the value $1$ or $0$ randomly with probabilities $p$ and $1-p$, respectively. However, the difference with the traditional ER model is that the choice of $A_{ij}$ is independent of $A_{ji}$ and, as a result, $\mathbf{A}$ is the non-symmetric matrical representation of a directed network where $pN(N-1)$ is the average number of links in this structure.
\\[2mm]
In Fig. \ref{Fig_9}, we explore three random networks with size $N=50$ generated with the value $p=0.04$, following the same approach presented for the analysis of regular topologies. We use connected networks for all the nodes to analyze the modifications in the eigenvalues introduced by the fractional transport. The eigenvalues of the transition probabilities and the respective networks are shown in Figs. \ref{Fig_9}(a)-(c). In this representation, for each value $\gamma$, we calculate numerically the fractional Laplacian matrix given by Eq. (\ref{LfracDef}) and in this way, we have the elements that define the transition matrix $\mathbf{W}^{(\gamma)}$, for which we calculate the eigenvalues $\lambda_l ^{(\gamma)}$. Due to the connectivity of the network, in the cases explored, the eigenvalue $\lambda_1^{(\gamma)}=1$ is unique for all $\gamma$. The remaining eigenvalues are represented as black dots in the complex plane for $\gamma=1$, and with different colors we show how the reduction of $\gamma$ concentrates the eigenvalues around the origin with $\lambda_l^{(\gamma)}=-1/(N-1)$ for $l=2,3,\ldots, N$ in the limit $\gamma\to 0$.
 \\[2mm]
 In addition to the eigenvalues, it is useful to quantify the capacity of the fractional random walker to explore the whole network. In this case, we use the global MFPT $\langle \mathrm{T}(\gamma)\rangle$ defined as
\begin{equation}\label{GlobalMFPT}
\langle \mathrm{T}(\gamma)\rangle=\frac{1}{N^2}\sum_{i=1}^N\sum_{j=1}^N
	\left\langle T_{ij}(\gamma)\right\rangle
\end{equation} 
that gives the average of the mean first passage time $\left\langle T_{ij}(\gamma)\right\rangle$ considering all the initial nodes $i$ and all the nodes target $j$. In Fig. \ref{Fig_9}(d), we show the results obtained with the numerical values of the eigenvalues and left and right eigenvectors of the transition matrix $\mathbf{W}^{(\gamma)}$ for $0<\gamma\leq 1$. In addition, in the inset in this figure, we include the values of the Kemeny's constant $\mathcal{T}(\gamma)=\sum_{l=2}^N \left(1-\lambda_l^{(\gamma)}\right)^{-1}$ that only includes information of the eigenvalues of $\mathbf{W}^{(\gamma)}$.
\\[2mm]
The results in Fig. \ref{Fig_9} illustrate the different cases that can appear in the fractional transport on directed structures for the same type of random network. In the information in Fig.  \ref{Fig_9}(d), the global MFPT $\langle \mathrm{T}(\gamma)\rangle$ shows that in some cases, like in the network (a),  the effect of $\gamma$ can improve the efficiency of the random walker to reach a target, in comparison with the normal random walk $\gamma\to 1$. In networks (b) and (c) we see two cases where the long-range dynamics increases the values of $\langle \mathrm{T}(\gamma)\rangle$ in comparison with the normal dynamics. In addition, the eigenvalues in (a)-(c) reveal different spectral properties with the change of $\gamma$. However, in contrast with the regular cases explored with circulant matrices, in this case the Kemeny's constant is not a good descriptor of the global activity in the fractional transport, revealing that in the networks analyzed the eigenvectors of $\mathbf{W}^{(\gamma)}$ contain important information for a global characterization of the dynamical process.

\section{Conclusions}
In this work, we presented a general approach to examining ergodic Markovian nonlocal random walks generated through the fractional Laplacian of directed networks. This formalism is explored for different types of strongly connected networks defined in terms of circulant matrices and also for random directed networks. We analyzed the eigenvalues of transition matrices that define the random walk and the effect of the fractional dynamics in the spectrum, showing how eigenvalues are modified in the complex plane. In addition, for regular networks, we analyze the Kemeny's constant, which gives a good description of the global dynamics. With this global time, defined only in terms of the eigenvalues, we analyzed the effect of biased nonlocal transport, showing that  the exploration of the network can be more effective in some particular cases reducing Kemeny's constant. We also identified different configurations for which the capacity of the fractional random walker to reach any node is reduced in comparison with a random walker with hops to neighbor nodes. This is a fundamental difference with the results observed in undirected networks where the fractional dynamics always improve the transport through long-range displacements. Finally, we explored the transport of random directed networks; in this case, the global activity is analyzed using the average of the mean first passage time to all the nodes considering all the initial conditions. This quantity depends on the eigenvalues and eigenvectors of the transition matrix that defines the process. Our findings and methods introduced are general and pave the way to further extensions for the exploration of fractional dynamical processes on directed structures with possible applications in the understanding of human mobility, data analysis, synchronization, among others.

\section{Appendix A: Properties of the Fractional Laplacian}
\label{AppendA}
In this appendix, we demonstrate briefly that the essential good properties (i)-(iii) of the Laplacian in Eq. (\ref{Laplacian}) are conserved by the fractional Laplacian matrix ${\mathbf L}^{\gamma}$ in the interval $0<\gamma \leq 1$  in order to guarantee that the fractional transition matrix (\ref{wijfrac}) is stochastic, i.e. 
$0\leq w_{i\to j}^{(\gamma)}\leq 1$ with $\sum_{l=1}^Nw_{i\to l}^{(\gamma)}=1$.
By the same demonstration we show that for the class of {\it strongly connected directed graphs} (i.e. between each pair of nodes $i,j$ at least one incoming and one outgoing connecting path of finite length exists $i \leftrightarrow j$ \cite{benzi2020fractional}) described by a Laplacian 
matrix of the form (\ref{Laplacian}) the fractional walk with transition matrix (\ref{wijfrac}) is 
{\it aperiodic ergodic}. 
To this end let us introduce $\Lambda > \max(k^{(\mathrm{out})}_i)$ thus the matrix 
$\Lambda \mathbb{I} -{\mathbf L}$ has uniquely non-negative entries $(\Lambda - k^{(\mathrm{out})}_i)\delta_{ij}+\Omega_{ij} \geq 0 $. 
Then it follows that $(\Lambda \mathbb{I} -\mathbf{L})^n$ and hence $e^{t (\Lambda {\mathbb I}-\mathbf{L})}= \sum_{n=0}^{\infty} \frac{t^n}{n!}(\Lambda {\mathbb I} -\mathbf{L})^n    =e^{\Lambda t} e^{-t\mathbf L} $ ($t>0$) also conserve this property (where for a strongly connected structure there exists a $n_0$ such that all entries $[(\Lambda {\mathbb I} -\mathbf{L})^n]_{ij} >0$ ($n \geq n_0$) are strictly
positive, for non-connected structures the inequality $[(\Lambda {\mathbb I} -\mathbf{L})^n]_{ij}  \geq 0$ remains true for all $n$.
Then for a strongly connected structure since $e^{\Lambda t} >0$ 
the matrix exponential $(e^{-t \mathbf L})_{ij} >0$. It follows then that the non-diagonal elements $(\mathbb{I} -  e^{-t\mathbf L})_{ij}  = - (e^{-t\mathbf L})_{ij} <0 $ ($i\neq j$) are uniquely negative (in a non-connected structure non-positive). Since the zero eigenvalue to the constant eigenvector $\langle i|\Psi_1\rangle =\frac{1}{\sqrt{N}}$ of ${\mathbf L}$ is conserved by the matrix function $(\mathbb{I} -  e^{-t\mathbf L})|\Psi_1\rangle = 0$,
it follows that
\begin{equation}
\label{it-holds}
(\mathbb{I} -  e^{-t\mathbf L})_{ii} = - \sum_{j\neq i}^N (\mathbb{I} -  e^{-t\mathbf L})_{ij} >0
\end{equation}
in a strongly connected directed graph (and $(\mathbb{I} -  e^{-t\mathbf L})_{ii} \geq 0$ if the network is not strongly connected).
Applying (\ref{mellin-repres}) on both sides of this relation yields Eq. (\ref{fracout}) and conserves the signs in Eq.  (\ref{it-holds}), 
i.e. $k_i^{(\gamma)}=(\mathbf{L}^\gamma)_{ii}  >0$ together with $(\mathbf{L}^\gamma)_{ij} <0 $ for $i\neq j$ in strongly connected directed graphs. If the graph is disconnected properties (i)-(iii) still hold; however the Laplacian matrix contains blocks of zero entries which are conserved by all integer powers of the Laplacian matrix and
hence also by the matrix exponential $e^{-t\mathbf L}$ leading by virtue of Eq. (\ref{mellin-repres}) that these blocks of zero-entries are still
conserved in the fractional Laplacian matrix, and hence in the resulting fractional walk ergodicity is lost. 
Compare especially the Laplacian and fractional Laplacian, respectively, in Figs. \ref{Fig_1}(a)-(b).
In Fig. \ref{Fig_1}(a) the block of zero entries of the non-strongly connected Laplacian matrix is conserved by the 
fractional Laplacian.
\\[2mm]
In this way we have demonstrated that the good
properties (i)-(iii) of ${\mathbf L}$ are indeed conserved by the fractional Laplacian matrix ${\mathbf L}^{\gamma}$ in the interval of convergence $0<\gamma <1$ of the integral representation in Eq. (\ref{mellin-repres}) where for strongly connected directed graphs
the fractional Laplacian generates an aperiodic ergodic walk. In this proof we did not make use of
whether or not the Laplacian matrix of a directed graph is diagonalizable. It includes therefore also the cases where the Laplacian matrix has Jordan canonical form.
For a more detailed analysis (however focused on undirected graphs) we refer to Ref. \cite{FractionalBook2019}.
\section{Appendix B: Fractional Laplacian for directed rings}
\label{AppendB}
In this Appendix, we analyze the elements of $\mathbf{L}^\gamma$ for a directed ring of finite size $N$ where $N$ is not necessarily large. Let us consider the Laplacian $\mathbf{L}$ for a directed ring defined with elements $c_0=c_{N-1}=1$ and $c_m=0$ for $m=1,2,N-2$.  Using Eq. (\ref{mu_circulant}) we have the eigenvalues
\begin{equation}
\label{Laplacianccirculant}
\mu_{\ell} = \sum_{m=1}^N c_m(1-e^{i\varphi_{\ell}m}),\qquad \varphi_{\ell}=\frac{2\pi}{N}(\ell-1).
\end{equation}
Therefore, the Laplacian eigenvalue is given by 
\begin{equation}
\label{mu-ell}
\mu_{\ell} =  1-e^{i\varphi_{\ell}(N-1)}=1-e^{-i\varphi_{\ell}}
\end{equation}
In addition, the elements of the fractional Laplacian matrix given by Eq. (\ref{Lijringdirected}) for $0<\gamma \leq 1$ are determined by
\begin{align}
\nonumber
(\mathbf{L}^{\gamma})_{pq} &= (\mathbf{L}^{\gamma})_{q-p} =\sum_{\ell=1}^N \mu_{\ell}^{\gamma} 
\langle p|\Psi_{\ell} \rangle\langle \Psi_{\ell}|q\rangle\\  \label{fraclap}
&=\frac{1}{N} \sum_{\ell=1}^N (1-e^{-i\varphi_{\ell}})^{\gamma} e^{i\varphi_{\ell}(q-p)}.
\end{align}
Now we can expand the fractional Laplacian eigenvalue $\mu_{\ell}^{\gamma}$ as (this series is converging)
\begin{align}
\nonumber
\mu_{\ell}^{\gamma} = 
(1-e^{-i\varphi_{\ell}})^{\gamma}&= \sum_{m=0}^{\infty} (-1)^m { \gamma \choose  m} e^{-i\varphi_{\ell} m}\\
&= \sum_{s=0}^{N-1} e^{-i\varphi_{\ell}s} {\cal A}_s^{(\gamma)}  \label{expansion}
\end{align}
with 
\begin{equation}
\label{this-res}
{\cal A}_s^{(\gamma)} = (-1)^{s+Nt} \sum_{t=0}^{\infty}{  \gamma \choose s+Nt}.
\end{equation}
In this result, we put $m=s+Nt$ and apply the $N$-periodicity condition $e^{-i\varphi_{\ell}m}=
e^{-i\varphi_{\ell}(s+Nt)} = e^{-i\varphi_{\ell}s}$ (with $e^{-i\varphi_{\ell}Nt} = e^{-2\pi i(\ell-1)t}=1$).
We see especially that Eq. (\ref{this-res}) indeed holds for finite $N \geq 2$.
\\[2mm]
Now, using the orthogonality property
\begin{equation}
\label{orythogona}
\frac{1}{N} \sum_{\ell=1}^N 
e^{i(q-p -s)\varphi_{\ell}} =\delta_{q-p,s} ,\qquad s=0,\ldots N-1,
\end{equation}
and combining Eqs. (\ref{mu-ell})-(\ref{orythogona}), we obtain the decomposition of the elements of the circulant fractional Laplacian matrix (\ref{fraclap}), i.e. the fractional Laplacian matrix for the finite ring in Eq. (\ref{Lijringdirected}), for $N$ finite but not necessarily large
\begin{align}
\label{elements} 
(\mathbf{L}^{\gamma})_{pq} &= \frac{1}{N} \sum_{\ell=1}^N (1-e^{-i\varphi_{\ell}})^{\gamma} e^{i\varphi_{\ell}(q-p)} \\ 
&= {\cal A}_{q-p}^{(\gamma)} = (-1)^{q-p+Nt} \sum_{t=0}^{\infty} { \gamma \choose q-p +Nt }
\end{align}
for $q-p=0,\ldots , N-1$. Hence we get 
\begin{multline} 
(\mathbf{L}^{\gamma})_{pq}= {\cal A}_{q-p}^{(\gamma)} = 
(-1)^{q-p} {\gamma \choose q-p }\\+ (-1)^{q-p+Nt} 
\sum_{t=1}^{\infty} 
{\gamma \choose q-p +Nt} ,\hspace{1cm} \gamma \in (0,1]
\end{multline}
(circulant). For the {\it finite directed ring} all matrix elements of the fractional Laplacian are non-vanishing, i.e. 
this structure hence is strongly connected and hence aperiodic ergodic.
We see in this result for the {\it finite ring} that the first term in the series ($t=0$), namely 
\begin{equation}
\label{fractriangle}
(\mathbf{L}^{\gamma})_{pq}^{(\infty)} = \left\{\begin{array}{l}
(-1)^{q-p} {\gamma \choose q-p}   ,\hspace{0.3cm} q-p \geq 0 \\ \\
0 ,\hspace{1cm} q -p < 0
\end{array}\right.
\end{equation}
is the matrix element of Eq. (\ref{FracLap_infdir_ring}) for the directed {\it infinite ring}. The fractional degree $(\mathbf{L}^{\gamma})_{qq}^{(\infty)}=1$ and $(\mathbf{L}^{\gamma})_{pq}^{(\infty)}<0$ for $q>p$ whereas 
$(\mathbf{L}^{\gamma})_{pq}^{(\infty)}=0$ is null for $q<p$. (\ref{fractriangle}) is an upper triangular circulant matrix where all
entries below the main diagonal are null.
\\[2mm]
The transition matrix for the infinite ring then writes with Eq. (\ref{wijfrac}) 
\begin{equation}
\label{transmatsibuya}
    w_{p\rightarrow q}^{(\gamma,\infty)} = \left\{\begin{array}{l} \displaystyle (-1)^{q-p-1} {\gamma \choose q-p} >0, \hspace{0.3cm} q-p >0 
     \\ \\ \displaystyle  0 ,\hspace{1cm} q-p \leq 0
    \end{array}\right. 
\end{equation}
for $\gamma \in (0,1)$ where the local limit $w_{p\rightarrow q}^{(1,\infty)}=\delta_{q-p,1}$ gives the deterministic walk where the walker in each step hops to its right-sided next neighbor node.
We observe that $w_{p\rightarrow q}^{(\gamma,\infty)}=0$ for $q\leq p$, i.e. for the infinite ring $N\rightarrow \infty$ the walker can only 
make jumps $p \to q$ such that $q - p > 0 $ with strictly increasing node numbers.
Conversely to the finite ring,
in the infinite ring limit no return path exists, and hence the infinite ring is not any more strongly connected 
thus ergodicity is lost. Indeed the fractional transition matrix (\ref{transmatsibuya}) of the infinite directed ring is an upper triangular circulant matrix where all elements above the main diagonal are strictly positive whereas all
entries below and in the main diagonal are null.
Eq. (\ref{transmatsibuya}) for $\gamma \in (0,1)$ can be identified with the transition probabilities of the {\it Sibuya walk} which is a strictly increasing walk on the positive integer line. The Sibuya walk is of utmost importance in models with power-law distributed (fat-tailed) long-range jumps. We refer to the recent article in Ref. \cite{PachonPolitoRicciuti2020} for a general outline and thorough analysis of properties (and see also the references therein). 
In Eq. (\ref{elements}) additionally to the infinite ring elements we have 
the image series $\sum_{t=1}^{\infty}(\ldots)$ where this additional contribution of the image terms
for $N$ finite but large can be (roughly) estimated as
\begin{equation}
\label{estimate}
\sum_{t=1}^{\infty} \frac{(Nt)^{-\gamma-1}}{\Gamma(-\gamma)} \approx \int_N^{\infty} \frac{\tau^{-\gamma-1}}{\Gamma(-\gamma)}{\rm d}\tau \sim -\frac{N^{-\gamma}}{\Gamma(1-\gamma)} .
\end{equation}
It follows that the infinite ring matrix elements (\ref{FracLap_infdir_ring}) already are a good approximation
for rings with $N$ large but not necessarily infinite.
In Ref. \cite{FractionalBook2019}, in section {\it 6.2.3. Fractional Laplacian of the finite ring}, we consider the fractional Laplacian of finite undirected rings and obtain analog results (see there Eqs. (6.27)-(6.31)) where we employ the same periodicity argument as here. 
%
%
%


%

\end{document}